\documentclass[reprint,superscriptaddress,amsmath,amssymb,aps,longbibliography
]{revtex4-2}
\usepackage{dcolumn, comment, amsmath, amsfonts, physics, mathcomp, bm, mathtools}
\usepackage{float, color}
\usepackage{ascmac, ulem}

\usepackage{placeins, tcolorbox}
\usepackage[dvipsnames]{xcolor}
\usepackage{txfonts}
\usepackage[colorlinks,citecolor=Violet,linkcolor=BrickRed]{hyperref}  

\begin{document}
\preprint{APS/123-QED}

\title{Topological transition induced by selective random defects on a honeycomb lattice}

\author{Sogen Ikegami}
\email{ikegami-sogen443@g.ecc.u-tokyo.ac.jp}
  \affiliation{Department of Applied Physics, The University of Tokyo, Tokyo 113-8656, Japan}
\author{Kiyu Fukui}%
  \affiliation{Department of Physical Sciences, Ritsumeikan University, Kusatsu, Shiga 525-8577, Japan}
\author{Shun Okumura}%
  \affiliation{Department of Applied Physics, The University of Tokyo, Tokyo 113-8656, Japan}
  \affiliation{Quantum-Phase Electronics Center (QPEC), The University of Tokyo, Hongo, Tokyo 113-8656, Japan }
  \affiliation{RIKEN Center for Emergent Matter Science, 2-1 Hirosawa, Wako 351-0198, Japan}
\author{Yasuyuki Kato}%
  \affiliation{Department of Applied Physics, University of Fukui, Fukui 910-8507, Japan}
\author{Yukitoshi Motome}%
  \email{motome@ap.t.u-tokyo.ac.jp}
  \affiliation{Department of Applied Physics, The University of Tokyo, Tokyo 113-8656, Japan}

\date{\today}

\begin{abstract}
We investigate how the spectral and topological properties of electron systems evolve 
on a lattice that interpolates between the honeycomb and its 1/6-depleted structures through the introduction of selective random defects.
We find that in certain parameter regimes, the topological properties of the two lattice systems are smoothly connected,
whereas in other regimes, selective random defects induce a topological transition.
Analysis based on an effective model
reveals that the effect of selective random defects can be understood 
as a modulation of hopping amplitudes.
Our results highlight the potential for designing and controlling 
the spectral and even topological properties of electronic systems across a wide range of material platforms.
\end{abstract}

\maketitle


\section{\label{sec1}Introduction}

  Topology has emerged as a central concept in modern condensed matter physics,
  offering a unifying framework for understanding diverse quantum phases beyond the conventional Landau-Ginzburg-Wilson paradigm
  based on spontaneous symmetry breaking.
  Topological phases exhibit striking phenomena, such as the quantum anomalous Hall effect,
  where the Hall conductance of a two-dimensional system is quantized to an integer value \cite{PhysRevLett.45.494, PhysRevLett.50.1395, PhysRevLett.49.405}.
  Such systems, now known as Chern insulators, are characterized 
  by integer-valued topological invariants called Chern numbers \cite{PhysRevLett.49.405, KOHMOTO1985343}.
  Topological band structures also arise in gapless systems,
  for example, simple tight-binding models on honeycomb and kagome lattices host Dirac cones
  in momentum space. 
  Notably, these lattices can be constructed from periodic site depletions of the triangular lattice, as shown in Fig.~\ref{fig:lattice},
  highlighting the potential of lattice engineering for realizing novel electronic phases.
  
  In practice, however, material fabrication inevitably introduces imperfections,
  making disorder an intrinsic feature of real systems.
  Such disorder is not merely detrimental 
  but can also be harnessed as a tool for lattice engineering,
  enabling the design of electronic properties inaccessible in perfectly ordered systems \cite{doi:10.34133/2019/4641739, PhysRevMaterials.5.044003, MISHRA2024100052}.
  Therefore, it is essential to understand how electron systems respond to disorder, 
  both for assessing their robustness and for exploring new avenues of property control.
  The interplay between disorder and topological phases has been extensively studied:
  topological phases are generally robust against disorder unless
  it closes the energy gap or breaks the protecting symmetries \cite{Q_Niu_1984, PhysRevB.31.3372}.
  Moreover, disorder can induce topologically nontrivial phases, 
  such as the topological Anderson insulator \cite{PhysRevLett.102.136806, PhysRevB.80.165316, PhysRevLett.103.196805},
  and drive topological phase transitions \cite{PhysRevLett.105.115501_2010, doi:10.1143/JPSJ.80.053703_2011, Zhang_2013, PhysRevLett.113.046802_2014, PhysRevB.92.085410_2015, PhysRevB.97.024204_2018, Imura_2018, PhysRevB.100.054108_2019, PhysRevB.100.235102_2019, Yang:20}.
  
  Recent advances in engineered platforms---including two-dimensional van der Waals materials \cite{vdW2_2013, vdW_layer_rev_2017},
  graphene nanostructures \cite{Bai2010, Zhang2017},
  photonic crystals \cite{PhysRevB.80.155103, Lu2014, Yang:20},
  and meta-materials \cite{metamate_review_2019}---have 
  enabled controlled realizations of various types of disorder \cite{doi:10.34133/2019/4641739, simonov2019, MISHRA2024100052}. 
  Among these, selective random disorder, where randomness is introduced in a 
  spatially or structurally controlled manner, is particularly intriguing.
  For instance, in bilayer systems, disorder may be introduced selectively in only one of the layers, 
  or vacancy engineering may be applied exclusively to specific sublattices or orbitals.
  Such selective randomness lies in between the clean and fully disordered limits and may lead to qualitatively new physics.
  Indeed, it has been reported that a Chern insulator with selective disorder on only one of two sublattices of the honeycomb lattice exhibits 
  remarkable robustness against disorder and hosts critical metallic states \cite{PhysRevB.92.085410_2015, PhysRevB.93.245414_2016, PhysRevResearch.1.033129_2019}.
  However, it remains unclear whether such selective disorder merely interpolates between topologically distinct clean limits, 
  or can induce genuine topological phase transitions.
  This motivates us to investigate how selective random defects
  mediate the evolution of the spectral and topological properties between the clean limits.

  In this paper, we consider a honeycomb lattice structure with selective random defects,
  in which defects are randomly introduced among the sites that would be removed in a 1/6-depleted honeycomb lattice.
  By introducing a parameter that quantifies the degree of defect incorporation,
  we continuously interpolate between the pristine honeycomb lattice and the 1/6-depleted structure,
  and investigate the spectral and topological properties of the prototypical Haldane model on these structures.
  For topological characterization, we adopt three topological invariants:
  the local Chern marker, the crosshair marker, and the Bott index,
  since reciprocal space is ill-defined in the presence of defects.
  Additionally, we also examine the energy spectra under periodic boundary conditions (PBC) and open boundary conditions (OBC) 
  to verify the presence of topologically protected edge states.
  As a result, we find two distinct routes of the interpolation of the spectral and topological nature.
  One is a smooth connection between the honeycomb and its 1/6-depleted counterpart,
  and the other involves a topological transition at a specific defect concentration.
  To gain further insight into the topological transition induced by selective random defects,
  we construct an effective model by assuming that the defects can be approximated by
  hopping-amplitude modulations.
  The effective model qualitatively reproduces the spectral features, 
  and even captures the topological transition,
  suggesting that the selective random defects can effectively be regarded 
  as hopping-amplitude modulations.
  Our results open the possibility of controlling physical properties via lattice engineering,
  which can be realized across various material platforms.

  The organization of this paper is as follows.
  In Sec.~\ref{sec2}, we introduce a model on the honeycomb lattice structure with selective random defects,
  and define the defect ratio which quantifies how much defects are introduced.
  In addition, we introduce three topological invariants used in this study.
  In Sec.~\ref{sec3}, we present the spectral and topological properties for the model,
  as well as the analysis of the effective model.
  Finally, we summarize our results and discuss future perspectives in Sec.~\ref{sec4}.

\section{\label{sec2}Model and method}
  \hypertarget{sec2}{}
  \subsection{Lattice with selective random defects}

  \begin{figure}[t]
    \includegraphics[width=\columnwidth]{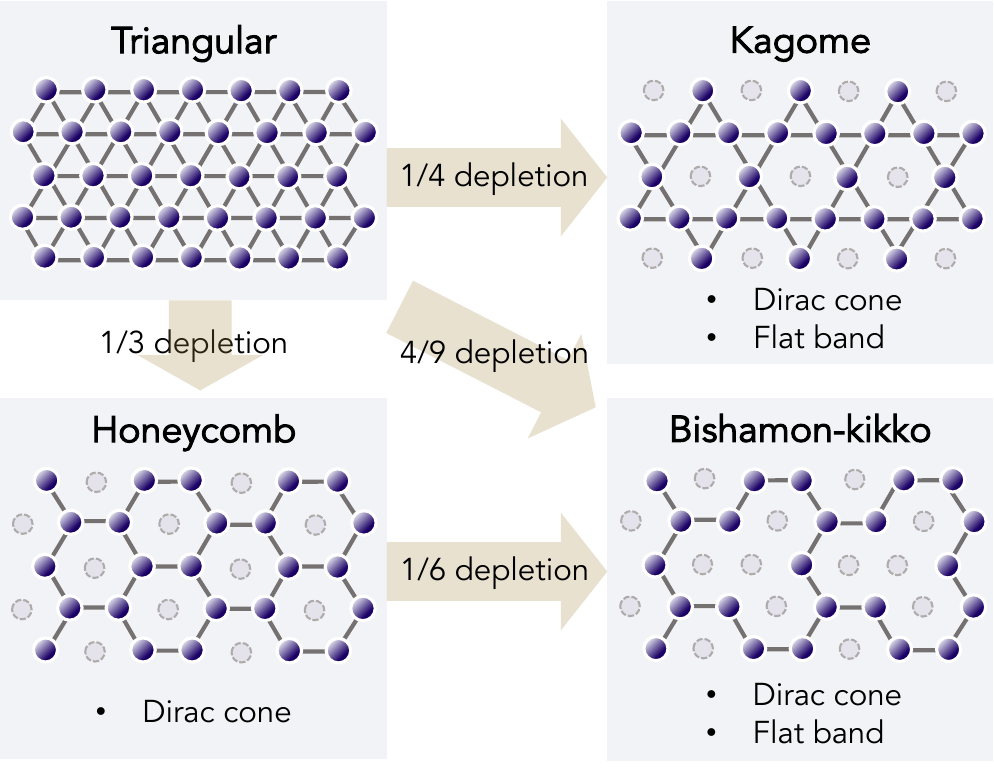}
    \caption{\label{fig:lattice} 
    Schematic pictures of the hierarchy of two dimensional lattices derived from a triangular lattice. 
    1/3 and 1/4 periodic site depletions applied to a triangular lattice lead to honeycomb and kagome lattices, respectively,
    which can host remarkable electronic band structures such as Dirac cones and flat bands. 
    Similarly, 1/6 periodic site depletion from the honeycomb lattice yields the Bishamon-kikko lattice, 
    which can exhibit both Dirac cones and flat bands in the electronic band structure.
    These examples illustrate that various periodic depletions applied to a single parent lattice can generate a wide variety of derived lattices 
    with rich electronic properties.}
  \end{figure}

  To investigate the effect of selective random defects on the topological nature of electrons, 
  we consider electron systems on two distinct lattices that satisfy the following condition: 
  (i) one lattice can be constructed from the other by the periodic site depletion,
  and (ii) electron systems on both lattices can host topologically nontrivial bands.
  For example, the honeycomb and kagome lattices can be obtained by periodically removing $1/3$ and $1/4$ of the sites from the triangular lattice, respectively, as illustrated in Fig.~\ref{fig:lattice}.
  Thus, the pairs of the triangular and the honeycomb lattices, and the triangular and the kagome lattices, satisfy the first condition.
  However, to construct a model that satisfies the second condition on the triangular lattice,
  we have to go beyond the simplest nearest-neighbor tight-binding model.
  In contrast, both the honeycomb and kagome lattices exhibit Dirac cones even in the nearest-neighbor tight-binding models,
  and are thus naturally suited for realizing the models that satisfy the second condition.
  On the other hand, it is impossible to transform the honeycomb and kagome lattices into each other through periodic site depletion (see Fig.~\ref{fig:lattice}).
  Therefore, the pair of the honeycomb and the kagome lattices meets only the second condition.
  Another pair can be generated by a $1/6$ site depletion from the honeycomb lattice, which gives rise to a $C_3$-symmetric lattice.
  This lattice belongs to a class of superhoneycomb systems \cite{PhysRevLett.71.4389}, 
  and is called the clover lattice \cite{doi:10.1126/sciadv.adg0028, chen2022emergent} or the Bishamon-kikko (BK) lattice \cite{PhysRevB.110.245107, ztny-z9yb}.
  The corresponding nearest-neighbor tight-binding model on the BK lattice has been reported to exhibit Dirac cones and flat bands \cite{doi:10.1126/sciadv.adg0028, chen2022emergent, PhysRevB.110.245107}.
  Thus, the honeycomb and BK lattices form a pair that satisfies the two conditions.
  In the following, we interpolate between these two lattices by selectively depleting sites, 
  and investigate how the topological properties of the electron systems evolve.

\begin{figure}[t]
    \includegraphics[width=0.9\columnwidth]{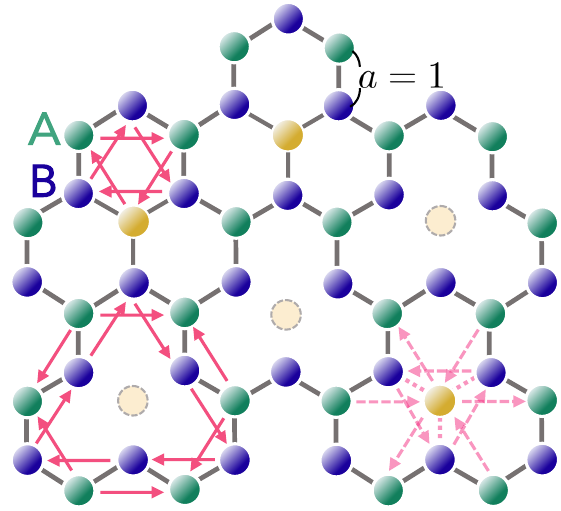}
    \caption{\label{fig:model} Schematic picture of the Haldane model on the honeycomb lattice with random depletion from selected sites, connecting to the BK lattice.
    The defect ratio $r$ is defined as the fraction of defect sites among the yellow-marked ones [Eq.~\eqref{eq:def-r}]. The figure represents an example with the defect ratio $r=3/6=0.5$. 
    The pink arrows indicate the direction of complex next-neighbor hopping with a phase $\phi$ in Eq.~\eqref{eq:Hamiltonian1}.
    The dashed translucent pink bonds and arrows represent hoppings that are modulated by the parameter $\alpha$ 
    in an effective model on the honeycomb lattice introduced in Sec.~\ref{sec3-3}}.
  \end{figure}

  Specifically, the BK lattice is constructed by depleting the yellow sites in Fig.~\ref{fig:model}.
  To connect the two lattices, we consider the ``selective random defects'', which means the random site depletion only from the yellow sites in Fig.~\ref{fig:model}.
  To quantitatively characterize how closely an obtained lattice structure resembles the honeycomb or the BK lattices, 
  we define the defect ratio $r$ as 
  \begin{equation}\label{eq:def-r}
    r = \frac{N_{\text{d}}}{N_{\text{hc}}-N_{\text{BK}}},
  \end{equation}
  where $N_{\text{hc}}$ and $N_{\text{BK}}$ represent the number of sites in the original honeycomb lattice and in the BK lattice obtained by periodically depleting the honeycomb lattice, respectively,
  and $N_{\text{d}}$ denotes the number of defect sites introduced into the original honeycomb lattice.
  For example, in Fig.~\ref{fig:model}, $r=3/6=0.5$, because $N_{\text{hc}}-N_{\text{BK}}$ is the number of yellow marked sites and equal to $6$,
  and $N_{\text{d}}=3$.
  We note that $r=0$ corresponds to the clean honeycomb lattice, while $r=1$ gives the BK lattice.
  
  To investigate the topological properties of electron systems on such lattice structures with selective random defects, 
  we adopt the prototypical Haldane model of spinless fermions \cite{PhysRevLett.61.2015}
  on those lattice structures.
  The Hamiltonian is given by
  \begin{equation}\label{eq:Hamiltonian1}
    \hat{H} = \sum_i M_i \hat{c}_i^\dagger \hat{c}_i + \left[ \sum_{\langle i,j \rangle} t_1\hat{c}_i^\dagger \hat{c}_j + \sum_{\langle\langle i,j \rangle\rangle} t_2e^{i\phi_{ij}}\hat{c}_i^\dagger \hat{c}_j+ \text{h.c.} \right],
  \end{equation}
  where $\hat{c}_i^\dagger$ and $\hat{c}_i$ are the creation and annihilation operators of spinless fermions at site $i$.
  $M_i$ represents a staggered site potential which takes the value $-M$ and $+M$ for A and B sublattices, respectively.
  In Fig.~\ref{fig:model}, the A sublattices are marked by the green and yellow, 
  while the blue sites indicate the B sublattices.
  $t_1$ and $t_2$ represent the transfer integrals,
  and the summations $\langle i,j \rangle$ and $\langle\langle i,j \rangle\rangle$ are taken for all nearest-neighbor pairs and second-neighbor pairs,
  represented by the gray bonds and the pink arrows in Fig.~\ref{fig:model}, respectively.
  $\phi_{ij}$ denotes the phase factor, taking the value $+\phi$ for $t_2$ hoppings in the clockwise direction with respect to the center of each plaquette, 
  as indicated by the pink arrows in Fig.~\ref{fig:model},
  and $-\phi$ for those in the counterclockwise direction.
  Hereafter, we fix $t_1=1$ as the energy unit, and the bond length between nearest-neighbor sites as $a=1$.

  \subsection{\label{sec2-2}Method}
  \hypertarget{sec2}{}

  \subsubsection{Energy spectra and density of states}
  To investigate the spectral properties of the Haldane model 
  on the lattice structures with selective random defects,
  we compute the energy spectrum and the density of states (DOS) by
  exact diagonalization of the single-particle Hamiltonian
  under both PBC and OBC.
  The DOS is obtained from the energy eigenvalues $\{\epsilon_i\}$ as
  \begin{equation}
    \text{DOS}(E) = \sum_{i} \delta(E - \epsilon_i).
  \end{equation}
  In actual calculations, we approximate the delta function by a Gaussian function with a width of $0.05$.
  By comparing the energy spectra under PBC and OBC, 
  we examine the presence or absence of edge states, which serve as signatures of 
  the topological character of the system.

  \subsubsection{Topological invariants in clean systems}
  Topological invariants of electron systems are often defined 
  as integrals of quantum geometrical quantities over closed manifolds.
  For example, the Chern number $C$ \cite{PhysRevLett.49.405, KOHMOTO1985343}, which characterizes the topology of band structures in a periodic lattice system, 
  is given by integrating the Berry curvature \cite{Berry1984} over the first Brillouin zone.
  To characterize the topological properties for the honeycomb lattice ($r=0$) and the BK lattice ($r=1$),
  we calculate the Chern number $C$ for each band by summing up the Berry curvature over the first Brillouin zone as
  \begin{equation}\label{eq:Chern}
    C = \frac{1}{2\pi} \int_{\text{1st BZ}} d\bm{k} \ \Omega_n(\bm{k}).
  \end{equation}
  Here, $\Omega_n(\bm{k}) = [\nabla_{\bm{k}} \times \bm{A}_{n\bm{k}}]_{k_z}$ is the $z$ component of the Berry curvature of the $n$-th band,
  where $\bm{A}_{n\bm{k}} = -i\expval{\nabla_{\bm{k}}}{u_{n\bm{k}}}$ denotes the Berry connection and
  $\ket{u_{n\bm{k}}}$ is the eigenstate in the 
  $n$-th band at momentum $\bm{k}$ with energy $\epsilon_n(\bm{k})$.
  In actual calculations of the Berry curvature and the Chern number,
  we adopt the Fukui-Hatsugai-Suzuki method, 
  in which we sum up the discretized Berry curvature over the first Brillouin zone \cite{doi:10.1143/JPSJ.74.1674}.

  \subsubsection{Topological invariants in disordered systems}
  In the systems with selective random defects ($0<r<1$), the translational symmetry is absent, 
  and the corresponding reciprocal space is no longer well-defined.
  To characterize and compute the topological properties of electrons in such disordered systems, 
  various methods have been proposed \cite{Corbae_2023}, such as 
  local markers \cite{KITAEV20062, PhysRevB.84.241106, PhysRevB.110.035146, PhysRevLett.129.277601, PhysRevB.106.155124, PhysRevB.107.045111, PhysRevB.109.014206},
  the Bott index \cite{Prodan_2010, Loring_2010, HASTINGS20111699, Prodan_2011, LORING2015383},
  and the spectral localizers \cite{LORING2015383, NoncommutGeom2020, PhysRevB.106.064109}.
  Among them, to verify the consistency of the calculation, we employ three different methods:
  the local Chern marker (LCM) \cite{PhysRevB.84.241106}, the crosshair marker \cite{PhysRevB.106.155124}, and the Bott index \cite{Prodan_2010, Loring_2010, HASTINGS20111699, Prodan_2011, LORING2015383}.
  For completeness, we briefly review the definitions and properties of the three markers.

  \paragraph{Local Chern marker}
    The LCM is a local quantity defined at a lattice site $\bm{r}_i$ \cite{PhysRevB.84.241106}.
    It is based on the idea of rewriting the Berry curvature defined in reciprocal space into real space quantity via inverse Fourier transformation.
    Several expressions have been reported for the LCM; 
    according to the formulation in Ref.~\cite{PhysRevB.95.121114}, it is defined by 
    \begin{equation}\label{eq:def-LCM1}
      \mathfrak{C}(\bm{r}_i) = -4\pi \Im \expval{\hat{\mathcal{P}}\hat{x}\hat{\mathcal{Q}}\hat{y}}{\bm{r}_i},
    \end{equation}
    where $\hat{x}$ and $\hat{y}$ denote the Cartesian components of the position operator $\hat{\bm{r}} = (\hat{x}, \hat{y})$, 
    each acting as an operator whose eigenvalues correspond to the spatial coordinates in the position basis.
    $\hat{\mathcal{P}}$ and $\hat{\mathcal{Q}}=\hat{1}-\hat{\mathcal{P}}$ are the projection operators to the occupied and unoccupied states, respectively.
    They are explicitly written as
    \begin{align}\label{eq:def-LCM2}
      \hat{\mathcal{P}} = \sum_{i:\epsilon_i < \mu} \ket{\psi_i}\bra{\psi_i}, \quad
      \hat{\mathcal{Q}} = \sum_{i:\epsilon_i > \mu} \ket{\psi_i}\bra{\psi_i},
    \end{align}
    where $\ket{\psi_i}$ is the $i$-th eigenstate of the Hamiltonian with eigenenergy $\epsilon_i$,
    and $\mu$ represents the chemical potential.

    The LCM applies to both clean and disordered systems, as well as to systems with PBC and OBC.
    For clean systems, the cell average of the LCM corresponds to the total Chern number of the occupied states as
    \begin{equation}\label{eq:def-LCM3}
      C_{\text{tot}} = \frac{1}{A_c} \sum_{i \in \text{cell}}\mathfrak{C}(\bm{r}_i),
    \end{equation}
    where the summation runs over the unit cell
    and $A_c$ represents the area per lattice site.
    In disordered cases, Eq.~\eqref{eq:def-LCM3} is replaced by an average over a macroscopic region of the system.

\begin{figure}[t]
      \hypertarget{fig:Real_space}{}
      \includegraphics[width=\columnwidth]{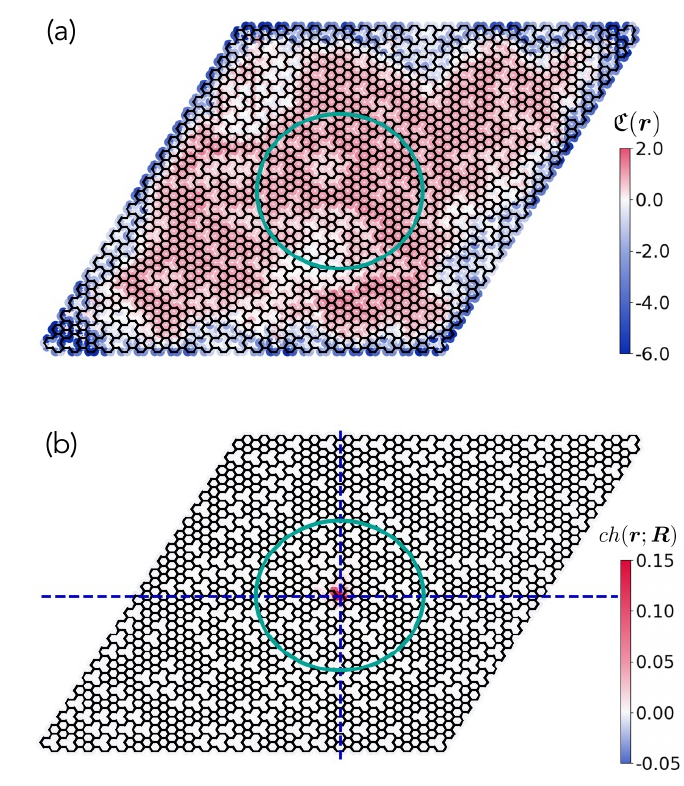}
      \caption{\label{fig:Real_space} A real space distribution of 
      (a) the local Chern marker and (b) the crosshair marker on a lattice under the OBC with the defect ratio $r=0.5$. 
      The Hamiltonian parameters are $M/t_2=-3$, $\phi=\pi/2$, and $t_2=0.3$.
      The intersection of the two blue dashed lines in (b) represents the position of the crosshair.
      To represent the bulk contributions, the LCM is averaged over the sites inside the green circle in (a),
      and the crosshair marker is summed over the same region in (b).
      }
    \end{figure}

    Figure \hyperlink{fig:Real_space}{\ref{fig:Real_space}(a)} displays an example of real space distribution of the LCM
    for the disordered system with defect ratio $r=0.5$ under the OBC. 
    As shown in this figure, the LCM takes positive values inside the sample
    and negative values near the boundaries.
    Consequently, the total LCM averages to zero over the entire sample due to the cancellation between 
    bulk and edge contribution.
    To extract the bulk contribution, instead of Eq.~\eqref{eq:def-LCM3},
    we calculate the average of the LCM within a central region of the system. 
    In the following calculations, we take an average over the sites within a circle of radius 15 from the center of the system, as
    \begin{equation}
      \overline{\mathfrak{C}} = \frac{1}{A_c} \sum_{i \in \text{circle}} \mathfrak{C}(\bm{r}_i),
    \end{equation}
    where the summation runs over the sites within the green circle in Fig.~\hyperlink{fig:Real_space}{\ref{fig:Real_space}(a)}.
    For the systems with defects,
    the area per site, $A_c$, is given by the following expression:
    \begin{equation}\label{eq:def-LCM4}
      A_c =  \frac{\frac{3\sqrt{3}}{2}N }{ 2N(1-\frac{r}{6}) } = \frac{9\sqrt{3}}{2(6-r)},
    \end{equation}
    where $N$ is the number of unit cells of the original honeycomb lattice.
    In the case of Fig.~\hyperlink{fig:Real_space}{\ref{fig:Real_space}(a)}, $\overline{\mathfrak{C}}$ 
    takes a value close to $+1$, 
    indicating that the electronic system is topologically nontrivial with a topological invariant of $+1$.

  \paragraph{Crosshair marker}
    The crosshair marker is also a locally defined marker
    which detects topological features of the electronic wave functions under the OBC \cite{PhysRevB.106.155124}.
    While the LCM is not straightforward to interpret physically, 
    the crosshair marker is defined on the basis of a physical picture. 
    Specifically, the marker measures the local contribution from the position $\bm{r}_i$
    to the local cross conductivity, which is the locally defined Hall conductivity, at the position of the crosshair $\bm{R}$.
    Its explicit definition is given as
    \begin{equation}\label{eq:def-crosshair1}
      ch(\bm{r}_i;\bm{R}) = 4\pi \Im \expval{\hat{\mathcal{P}} \hat{\vartheta}_{R_x} \hat{\mathcal{P}} \hat{\vartheta}_{R_y} \hat{\mathcal{P}} }{\bm{r}_i}.
    \end{equation}
    The operators $\hat{\vartheta}_{R_{x,y}}$ in Eq.~\eqref{eq:def-crosshair1} act as projection operators onto one side of the system $x>R_x$ and $y>R_y$,
    defined using step functions $\theta(\cdot)$ as
    \begin{equation}\label{eq:def-crosshair2}
      \hat{\vartheta}_{R_{x}} = \sum_{i} \theta(x_i - R_x) \ket{\bm{r}_i}\bra{\bm{r}_i},
    \end{equation}
    and similarly for $\hat{\vartheta}_{R_y}$. 
    The summation in Eq.~\eqref{eq:def-crosshair2} runs over all sites $i$ 
    and $x_i$ represents the $x$ coordinate of site $i$.
    It has been analytically shown that the integration of the crosshair marker for the position of the crosshair $\bm{R}$
    over the whole sample is equal to the LCM, expressed as
    \begin{equation}\label{eq:def-crosshair3}
      \mathfrak{C}(\bm{r}_i) = \int_{\mathrm{sample}} d^2\bm{R} \ ch(\bm{r}_i;\bm{R}).
    \end{equation}

    Figure \hyperlink{fig:Real_space}{\ref{fig:Real_space}(b)} represents an example of 
    real space distribution of the crosshair marker on a lattice with defect ratio $r=0.5$.
    The intersection of the two blue dashed lines indicates the position of the crosshair $\bm{R}$.
    As shown in Fig.~\hyperlink{fig:Real_space}{\ref{fig:Real_space}(b)}, 
    the crosshair marker typically exhibits nonzero values near the crosshair, separating the bulk and edge contributions
    to the local Hall conductance at the crosshair position $\bm{R}$. 
    In this study, similar to the LCM, we fix $\bm{R}$ at the center of the system,
    and calculate the summation of the crosshair marker within a circle of radius 15 from the position of the crosshair $\bm{R}$ as
    \begin{equation}
      \overline{ch} = \sum_{i \in \text{circle}} ch(\bm{r}_i; \bm{R}).
    \end{equation}
    where the summation runs over the sites within the green circle in Fig.~\hyperlink{fig:Real_space}{\ref{fig:Real_space}(b)}.

  \paragraph{Bott index}
    The Bott index is a nonlocal topological invariant defined for finite-size systems,
    formulated within the framework of $K$-theory \cite{Prodan_2010, Loring_2010, HASTINGS20111699, Prodan_2011, LORING2015383}.
    It is defined as 
    \begin{equation}\label{eq:def-Bott1}
      B = \frac{1}{2\pi}\Im \Tr \log(\hat{V}\hat{U}\hat{V}^\dagger \hat{U}^\dagger),
    \end{equation}
    where $U = \hat{\mathcal{P}} e^{2\pi i \hat{X}} \hat{\mathcal{P}}$ and $V = \hat{\mathcal{P}} e^{2\pi i \hat{Y}} \hat{\mathcal{P}}$
    are projected position operators;
    $\hat{X}$ and $\hat{Y}$ represent the rescaled coordinate operators whose eigenvalues are normalized to the interval $[0,1)$,
    and $\hat{\mathcal{P}}$ is defined in Eq.~\eqref{eq:def-LCM2}.
    While the position operators are ill-defined under the PBC,
    $e^{2\pi i \hat{X}}$ and $e^{2\pi i \hat{Y}}$ are well-defined thanks to their periodicity.
    The Bott index quantifies the noncommutativity of these two projected position operators.
    In the thermodynamic limit, it becomes equivalent to the total Chern number of the occupied states \cite{Toniolo2022}.
  
  \subsubsection{Calculation conditions}
    The DOS and the three topological invariants are averaged over 60 independent lattice realizations, 
    each generated by introducing selective random defects with a specific value $r$.
    The number of sites used in the DOS and energy spectrum calculations ranges from $N_{\text{hc}}=10086$ 
    for the honeycomb lattice ($r=0$) to $N_{\text{BK}}=8405$ for the BK lattice ($r=1$).
    The calculations of the topological invariants are performed under the OBC, 
    and the number of sites of the system ranges from 3750 for the honeycomb lattice ($r=0$) to 3125 for the BK lattice ($r=1$).

\section{\label{sec3}Result}

  \subsection{\label{sec3-1}Smooth connection via selective random defects}
\begin{figure}[t]
      \hypertarget{fig:pattern1}{}
      \includegraphics[width=\columnwidth]{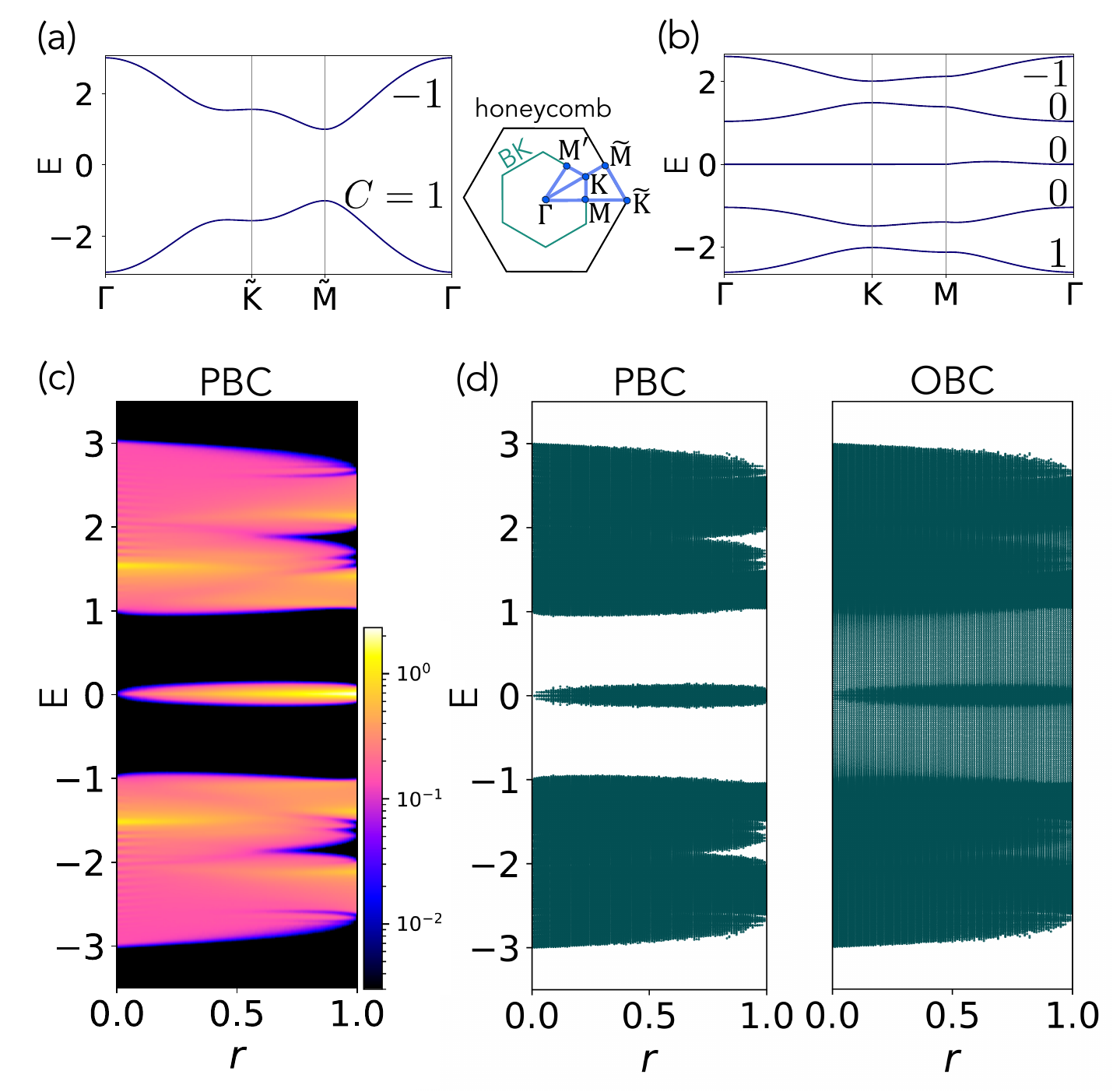}
      \caption{\label{fig:pattern1} Spectral properties of the Haldane model in Eq.~\eqref{eq:Hamiltonian1} at $M=0$, $t_2=0.3$, and $\phi=\pi/2$. 
      (a), (b) Band structures along the high symmetric lines in the Brillouin zones for the model on (a) the honeycomb lattice ($r=0$) and (b) the BK lattice ($r=1$). 
      The Chern numbers $C$ associated with each band are also shown.
      The inset represents the first Brillouin zone for the honeycomb (outer black hexagon) and BK (inner green hexagon) lattices.
      (c) DOS under the PBC averaged over 60 selective random defect configurations and 
      (d) energy spectrum under PBC and OBC as functions of the defect ratio $r$.}
    \end{figure}

    We first outline the general electronic structures of the Haldane model on the honeycomb and BK lattices,
    which correspond to the two limits of the defect ratio $r=0$ and $r=1$, respectively.
    The honeycomb lattice hosts two bands with Dirac cones at the $\tilde{\text{K}}$ point [see the inset of Fig.~\hyperlink{fig:pattern1}{\ref{fig:pattern1}(a)}],
    where topological transitions occur upon varying the model parameters such as $t_2$, $M$, and $\phi$.
    In contrast, the BK lattice consists of five sublattices, giving rise to five bands in the Brillouin zone,
    including flat bands depending on the parameters \cite{PhysRevB.110.245107}.
    As a representative case, we first take the model parameters as $M=0$, $t_2=0.3$, and $\phi=\pi/2$,
    and investigate the electronic states of the Haldane model on the lattice structure with selective random defects.
    We begin with the lattice with $r=0$ and $r=1$, i.e., the honeycomb and BK lattices respectively,
    in which the band structures are well defined due to the absence of randomness.
    Figures \hyperlink{fig:pattern1}{\ref{fig:pattern1}(a)} and \hyperlink{fig:pattern1}{\ref{fig:pattern1}(b)} show the band structures
    along the high-symmetric lines in the Brillouin zones, illustrated by the hexagons in the inset.
    The corresponding Chern numbers $C$ for each band are also indicated for both cases.
    Due to the absence of the staggered site potential, the A and B sublattices are equivalent,
    and thus the band structures preserve particle-hole symmetry.

    When $0<r<1$, as momentum is no longer a good quantum number, 
    we focus on the DOS and energy spectra instead of the band structures.
    Figure~\hyperlink{fig:pattern1}{\ref{fig:pattern1}(c)} shows the DOS under the PBC as a function of the defect ratio $r$,
    which is obtained by averaging over 60 selective random defect configurations for each $r$.
    The left panel of Fig.~\hyperlink{fig:pattern1}{\ref{fig:pattern1}(d)} displays the corresponding energy spectra under the PBC
    calculated for a single realization of selective random defect configurations.
    In the $r=0$ limit, a single bulk gap is present in the energy range of $-1<E<1$, 
    which immediately splits by the introduction of $r$ and smoothly evolves into two separate gaps in the $r=1$ limit.
    Furthermore, the spectra on both sides of this gap begin to split at $r \simeq 0.7$, 
    giving rise to two additional gaps at $E \simeq \pm 2$, so that four distinct gaps are present in total at $r=1$.
    Despite this reconstruction of the gap structure, 
    the top and bottom parts of the DOS and spectra appear to be smoothly connected between $r=0$ and $r=1$.
    Indeed, in both $r=0$ and $r=1$ limits, the top and bottom bands carry the Chern numbers $C=\pm1$, 
    as indicated in Figs.~\hyperlink{fig:pattern1}{\ref{fig:pattern1}(a)} and \hyperlink{fig:pattern1}{\ref{fig:pattern1}(b)}.
    This smooth evolution of the energy spectra between $r=0$ and $r=1$ suggests that 
    the topological character of the upper and lower bands remains intact for any $0<r<1$.

    To further confirm the topological nature, we compare the spectra under the OBC with those under the PBC.
    The right panel of Fig.~\hyperlink{fig:pattern1}{\ref{fig:pattern1}(d)} shows the energy spectra under the OBC, 
    presented in the same way as in the left panel for the PBC.
    As shown in the right panel, all bulk gaps obtained under the PBC are filled with energy levels under the OBC.
    These additional states originate from the open boundaries of the system,
    indicating the presence of topologically protected edge states.
    This is a hallmark of a topological phase, where energy gaps 
    in the bulk spectrum host localized edge modes under the OBC. 
    This topological property of the energy spectra is consistent with that of the band structures 
    in the limits of $r=0$ (honeycomb lattice) and $r=1$ (BK lattice), 
    where the top and bottom bands carry the Chern numbers $C=\pm1$:
    all the gapped states are Chern insulators with $C_{\text{tot}}=1$.
    Indeed, the three topological invariants, the LCM [Eq.~\eqref{eq:def-LCM1}], 
    the crosshair marker [Eq.~\eqref{eq:def-crosshair1}], and the Bott index [Eq.~\eqref{eq:def-Bott1}],
    take the value $+1$ for all $0\leq r \leq 1$
    when the chemical potential is set within the gap at $E \simeq \pm0.5$.
    Thus, the $r$ dependencies of the energy spectra reveal nontrivial topology in the presence of 
    selective random site depletion smoothly connecting two periodic systems for the present parameter set.
    
  \subsection{\label{sec3-2} Topological transition induced by selective random defects}

    \begin{figure}[thp]
    \hypertarget{fig:pattern2}{}
    \includegraphics[width=\columnwidth]{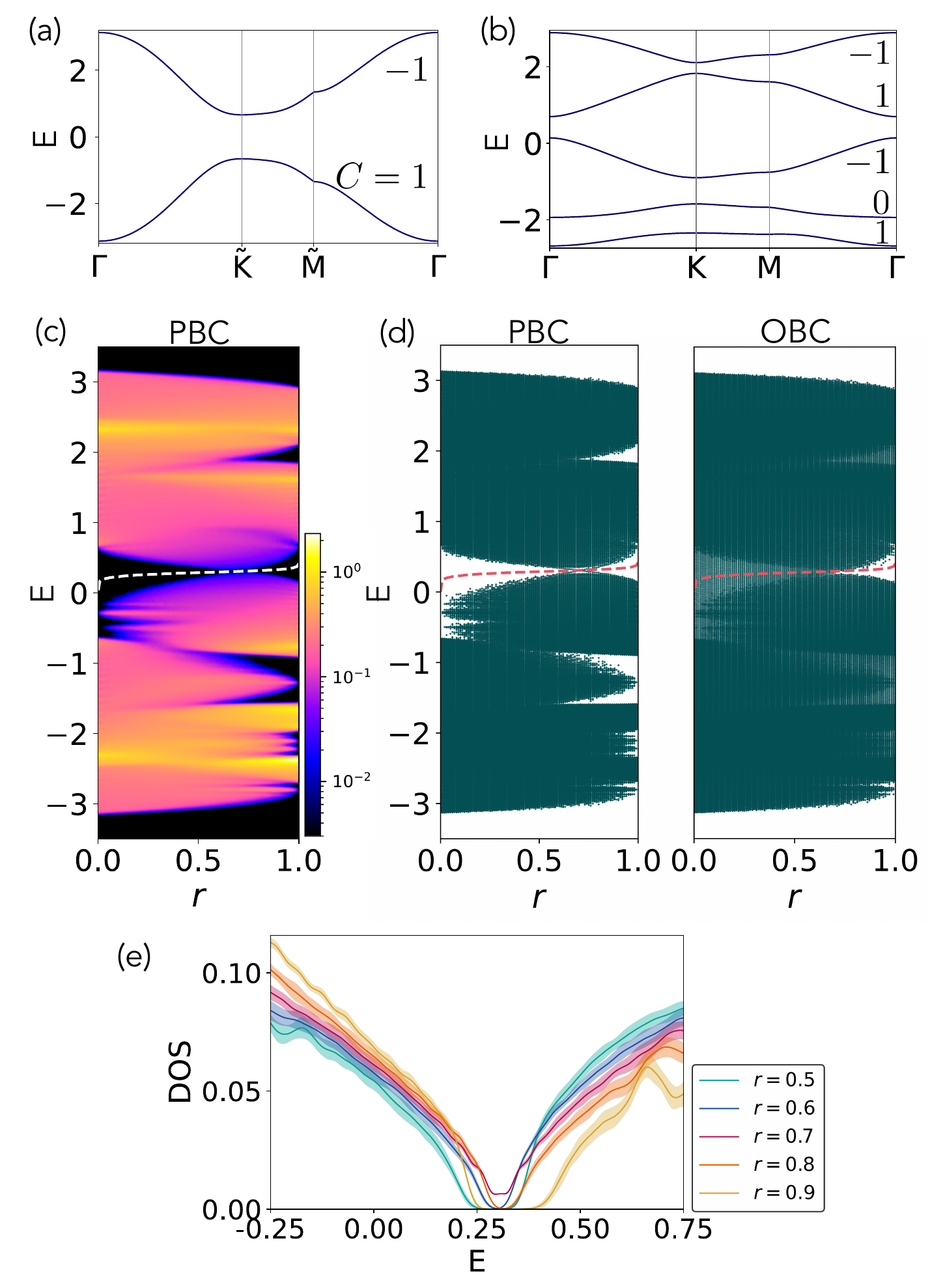}
    \caption{\label{fig:pattern2} Spectral properties of the Haldane model at $M/t_2=-3, t_2=0.3$, and $\phi=\pi/2$. 
    Band structures and Chern numbers $C$ of the Haldane model on (a) the honeycomb and (b) the BK lattices. 
    (c) DOS under the PBC and (d) energy spectrum under PBC and OBC 
    as functions of the defect ratio $r$. 
    (e) DOS at defect ratio $0.5, 0.6, 0.7, 0.8,$ and $0.9$ under the PBC.
    The solid lines represent the DOS averaged over 60 independent selective random defect configurations, and the shades indicate the standard deviation around each average.
    The white dashed line in (c) and red dashed lines in (d) represent 
    the chemical potential for calculating the topological invariants in Fig.~\ref{fig:marker}. }
    \end{figure}

    \begin{figure}[t]
      \includegraphics[width=0.9\columnwidth]{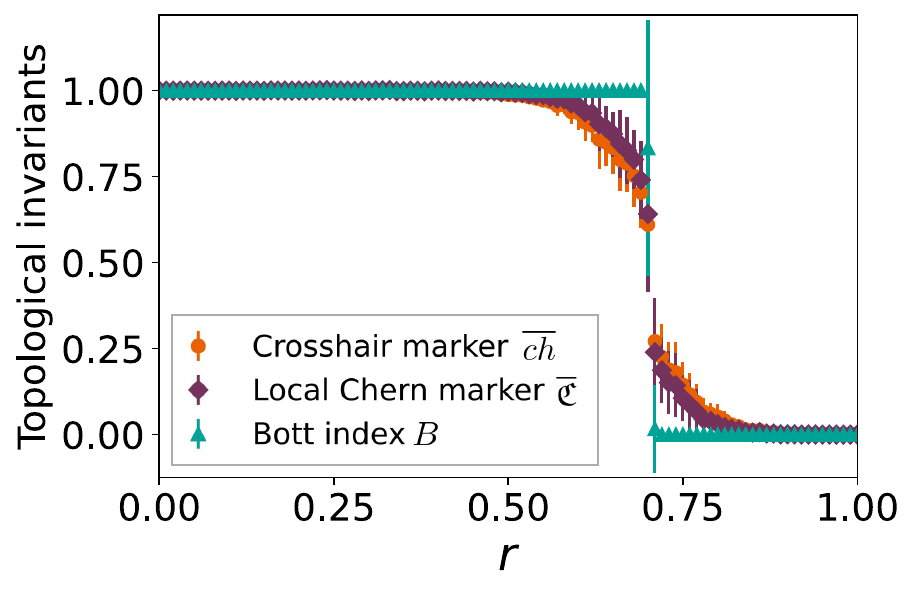}
      \caption{\label{fig:marker} $r$ dependencies of three topological invariants for the same model parameters as Fig.~\ref{fig:pattern2}. 
      The chemical potential is set within the relevant spectral gap, 
      as indicated by the white and red dashed lines
      in Figs.~\protect\hyperlink{fig:pattern2}{\ref{fig:pattern2}(c)} 
      and \protect\hyperlink{fig:pattern2}{\ref{fig:pattern2}(d)}. }
    \end{figure}
    Next we present the results for $M/t_2=-3$, $t_2=0.3$, and $\phi = \pi/2$.
    Figures \hyperlink{fig:pattern2}{\ref{fig:pattern2}(a)-\ref{fig:pattern2}(d)} show,
    in the same manner as Fig.~\ref{fig:pattern1},
    the band structures and the Chern numbers $C$ of each band for the honeycomb and BK lattices [Figs.~\hyperlink{fig:pattern2}{\ref{fig:pattern2}(a)} and \hyperlink{fig:pattern2}{\ref{fig:pattern2}(b)}],
    the DOS under the PBC [Fig.~\hyperlink{fig:pattern2}{\ref{fig:pattern2}(c)}], 
    and the energy spectra under both PBC and OBC [Fig.~\hyperlink{fig:pattern2}{\ref{fig:pattern2}(d)}].
    Due to the nonzero onsite staggered potential, 
    the band structures of the BK lattice and the energy spectra for $r>0$ do not preserve the particle-hole symmetry.
    
    At first glance, the DOS and the energy spectra appear smoothly connected between $r=0$ and $r=1$,
    as in the previous cases shown in Fig.~\hyperlink{fig:pattern1}{\ref{fig:pattern1}}.
    However, focusing on the energy gap around $E\simeq 0.3$, 
    it gradually narrows from $r=0$, 
    closes near $r\simeq 0.7$,
    and reopens as $r$ increases further.
    To provide a closer look at the DOS evolution across the gap-closing point,
    Fig.~\hyperlink{fig:pattern2}{\ref{fig:pattern2}(e)} presents the DOS as a function of the energy
    for the defect ratios $r=0.5$, $0.6$, $0.7$, $0.8$, and $0.9$.
    The solid lines in each color represent the DOS averaged over 60 independent selective random defect configurations,
    and the shades indicate the standard deviation around each average.
    Notably, as shown by the red line in Fig.~\hyperlink{fig:pattern2}{\ref{fig:pattern2}(e)}, 
    the DOS at $r=0.7$ exhibits a V-shape around $E \simeq 0.3$, similarly to that for the system with a Dirac cone.
    Furthermore, in the energy spectra under the OBC [the right panel of Fig.~\hyperlink{fig:pattern2}{\ref{fig:pattern2}(d)}], 
    energy levels appear inside the gap for $r\lesssim 0.7$,
    whereas such edge states are absent for $r\gtrsim 0.7$.
    These observations indicate that this spectral change with gap closing
    corresponds to a topological transition induced by selective random defects.

    To further confirm the topological transition, we calculate the topological invariants and directly examine their behavior.
    Figure \ref{fig:marker} shows the defect ratio $r$ dependence of three topological invariants: the LCM [Eq.~\eqref{eq:def-LCM1}], 
    the crosshair marker [Eq.~\eqref{eq:def-crosshair1}], and the Bott index [Eq.~\eqref{eq:def-Bott1}].
    The chemical potential is set in the middle of the relevant gap, 
    as indicated by the white and red dashed lines in Figs.~\hyperlink{fig:pattern2}{\ref{fig:pattern2}(c)} and \hyperlink{fig:pattern2}{\ref{fig:pattern2}(d)}, respectively.
    The symbols indicate the averaged values over 60 independent lattice systems with selective random defect configurations,
    and the error bars represent the standard deviation.
    Examples of real-space distributions of the LCM and the crosshair marker 
    for a lattice realization with $r=0.5$
    are shown in Fig.~\hyperlink{fig:Real_space}{\ref{fig:Real_space}(a)} and \hyperlink{fig:Real_space}{\ref{fig:Real_space}(b)}, respectively.
    We observe a clear jump in all topological invariants at $r\simeq0.7$,
    confirming that the spectral change with gap closing seen in Fig.~\ref{fig:pattern2}
    corresponds to a topological transition.
    Importantly, the two phases on either side of the jump, each characterized by distinct topological invariants,
    are connected to the clean honeycomb or BK systems without further gap closing.
    These results signify a genuine topological transition driven by selective random defects,
    rather than a topology breakdown caused by randomness,
    where sample dependencies smear out the sharp jump and lead to smooth changes.
    The insensitivity of the topological transition to defect configurations
    strongly suggests the presence of an underlying physical mechanism.
    To elucidate this phenomenon, we construct an effective model in the next section.    

  \subsection{\label{sec3-3} Effective model analysis of the topological transition}

    To capture the essence of selective random defects while allowing for tractable analysis,
    we construct a periodic effective model on the honeycomb lattice,
    in which selective random defects are represented as an effective modulation of the hopping amplitudes.
    This construction provides an approximate, coarse-grained description:
    instead of explicitly removing lattice sites, we incorporate the reduction of connectivity induced by defects
    into a uniform modulation of the hoppings.
    The Hamiltonian of the effective model is given by
    \begin{align}\label{eq:Heff}
      \hat{H}_{\text{eff}} &= \sum_i M_i \hat{c}_i^\dagger \hat{c}_i \notag \\
      &+ \left[\sum_{\langle i,j \rangle} \alpha_{ij}t_1 \hat{c}_i^\dagger \hat{c}_j + \sum_{\langle\langle i,j \rangle\rangle} \alpha_{ij}t_2 e^{i\phi_{ij}} \hat{c}_i^\dagger \hat{c}_j + \text{h.c.}\right],
    \end{align}
    where $\alpha_{ij} = \alpha \ (0\leq \alpha \leq 1)$ when the hopping process involves a defect site on the BK lattice, 
    as illustrated by the dashed translucent pink arrows and bonds in Fig.~\ref{fig:model}, and $\alpha_{ij}=1$ otherwise.
    The other parameters and notations are the same as those in Eq.~\eqref{eq:Hamiltonian1}.
    This model reduces to the original Haldane model on the honeycomb lattice when $\alpha=1$.
    In contrast, $\alpha=0$ corresponds to the Haldane model on the BK lattice
    with additional isolated sites at the centers of the plaquettes.
    The partially reduced hopping amplitudes with $\alpha$ are introduced to mimic 
    the impact of selective random defects.
    The band structures and topological phase diagrams of this model were studied in Ref.~\cite{PhysRevB.110.245107}.
  
    \begin{figure}[t]
      \includegraphics[width=0.95\columnwidth]{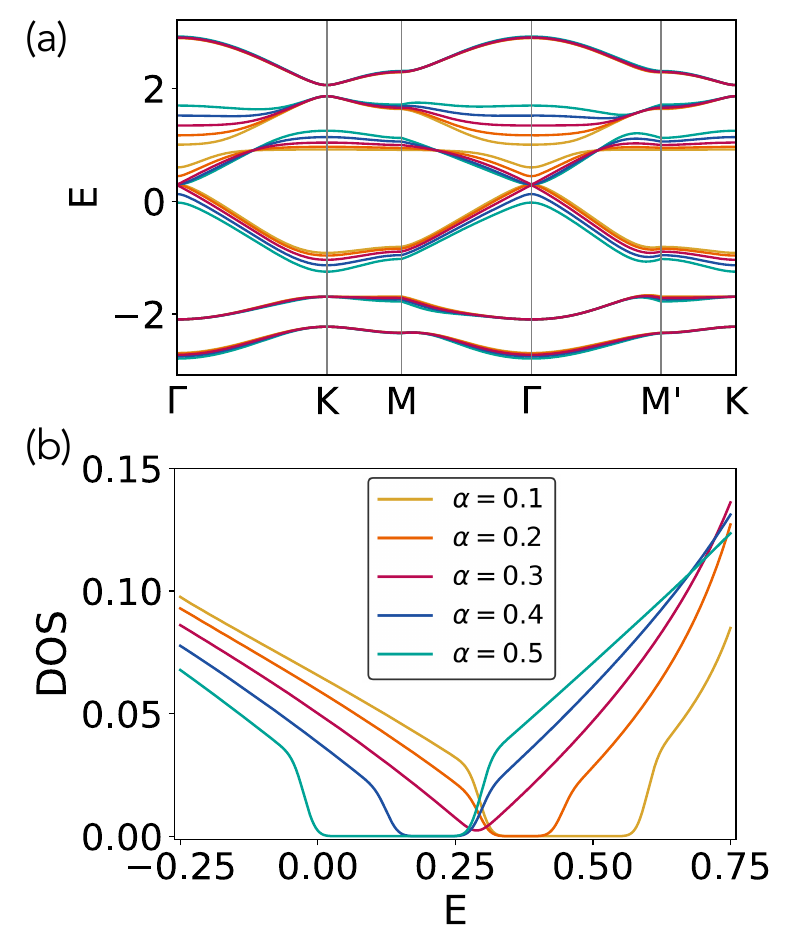}
      \hypertarget{fig:BKH_spec}{}
      \caption{\label{fig:BKH_spec} Spectral properties of the effective model [Eq.~\eqref{eq:Heff}] 
      varying $\alpha$ while fixing other parameters as $M/t_2=-3$, $t_2=0.3$, and $\phi=\pi/2$. 
      (a) Band structures along the high symmetric lines. 
      (b) DOS around the energy where band gap closes at $\alpha=0.3$.}
    \end{figure}
    
    For discussing the topological transition in Sec.~\ref{sec3-2},
    we employ the same model parameters, $M/t_2=-3$, $t_2=0.3$, and $\phi=\pi/2$.
    Figure \ref{fig:BKH_spec} displays the band structures and the DOS of the effective model
    for $\alpha=0.1$, $0.2$, $0.3$, $0.4$, and $0.5$.
    At $\alpha=0.3$, the band structure exhibits a linear band crossing at the $\Gamma$ point,
    as shown by the red curve in Fig.~\hyperlink{fig:BKH_spec}{\ref{fig:BKH_spec}(a)}.
    Correspondingly, the DOS displays a linear energy dependence around $E\simeq0.3$
    in Fig.~\hyperlink{fig:BKH_spec}{\ref{fig:BKH_spec}(b)}.
    For $\alpha \neq 0.3$, the Dirac node at the $\Gamma$ point is gapped out.
    These behaviors coincide with those observed through the topological transition induced by selective random defects at $r=0.7$ in Fig.~\hyperlink{fig:pattern2}{\ref{fig:pattern2}(e)}.
    This observation suggests that
    the effective model qualitatively captures the spectral characteristics of the Haldane model 
    on the lattices with selective random defects in this energy range.

    Furthermore, to investigate the topological characteristics at $\alpha=0.3$,
    we compute the Berry curvature density $D_{\Omega}(E)$, defined as
    \begin{equation}\label{eq:Domega}
      D_\Omega(E) = \sum_{n} \int_{\text{1st BZ}}\frac{d^2\bm{k}}{2\pi} \ \Omega_n(\bm{k}) \delta(E-\epsilon_n(\bm{k})),
    \end{equation}
    and its integrated value up to the energy $E$,
    \begin{equation}\label{eq:Cocc}
      C_{\text{occ}}(E) = \int_{-\infty}^{E} dE^\prime \ D_\Omega(E^\prime).
    \end{equation}
    The summation in Eq.~\eqref{eq:Domega} runs over all bands, and the integration
    is performed over the first Brillouin zone.
    $\epsilon_n(\bm{k})$ denotes the energy of the $n$-th band at momentum $\bm{k}$.
    In actual calculations, the delta function in Eq.~\eqref{eq:Domega} is approximated by a Gaussian function with a width of $0.05$
    as the same manner as the DOS calculation.
    When $E$ lies within an energy gap, $C_{\text{occ}}(E)$ corresponds to the sum of the Chern numbers 
    of all bands below the gap.

    \begin{figure}[t]
      \includegraphics[width=0.95\columnwidth]{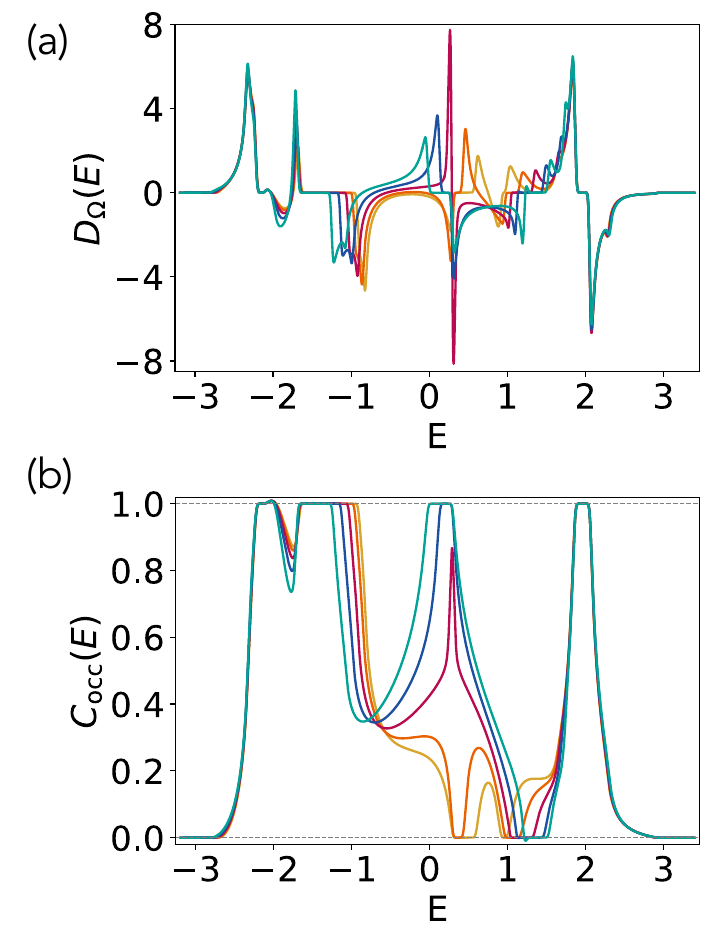}
      \hypertarget{fig:BKH_berry}{}
      \caption{\label{fig:BKH_berry} Energy dependencies of (a) the Berry curvature density and (b) the cumulative Berry curvature
      of the effective model in Eq.~\eqref{eq:Heff}. The model parameters and corresponding color codes are common to those in Fig.~\ref{fig:BKH_spec}.  }
    \end{figure}

    Figure \ref{fig:BKH_berry} shows the energy dependencies of the Berry curvature density $D_\Omega$
    and the cumulative Berry curvature, $C_{\text{occ}}$.
    Distinct peaks in $D_\Omega$
    and quantized plateaus in $C_{\text{occ}}$ 
    are observed around $E\simeq0.3$.
    For $\alpha=0.1$ and $0.2$ (yellow and orange curves),
    the plateau value in $C_{\text{occ}}$ is 0, while for $\alpha=0.4$ and $0.5$ (blue and green curves),
    it is quantized to $+1$. 
    This behavior indicates that the topological transition occurs at $\alpha=0.3$,
    where the Dirac node appears at the $\Gamma$ point in the band structure.
    We note that topological transitions do not occur
    within the effective model with parameters used in Sec.~\ref{sec3-1}, $M/t_2=0$, $t_2=0.3$, and $\phi=\pi/2$,
    corresponding to the smooth connection of the topological properties.
    These results demonstrate that the effective model qualitatively reproduces 
    the spectral and topological properties of the Haldane model on lattices with selective random defects.
    In particular, they suggest that the topological transition induced by selective random defects
    can be understood in terms of hopping-amplitude modulations.
    For the above parameter set, the topological transition occurs at $r\simeq 0.7$ for the disordered model in Eq.~\eqref{eq:Hamiltonian1} 
    and at $\alpha\simeq 0.3$ in the effective model in Eq.~\eqref{eq:Heff}. 
    For another parameter set, we show that the transition takes place at $r\simeq 0.5$ and $\alpha\simeq 0.5$ (see Appendix A).
    These observations suggest a quantitative relationship: 
    the topological transition induced by selective random defects with defect ratio $r$
    is described by that in the effective model with $\alpha \simeq 1 - r$.
    A more systematic investigation of this correspondence would be an interesting direction for future work.

\section{\label{sec4}Summary and perspectives}

  In summary, we have investigated the effects of selective random defects on the spectral and topological properties of an electronic system,
  focusing on the effects of such defects on the Haldane model on the honeycomb lattice.
  Depending on the model parameters, 
  we unveiled two distinct behaviors while varying the concentration of selective random defects: 
  a smooth connection between two clean limits
  and a transition between topologically different states.
  To further understand mechanism behind this topological transition, 
  we constructed an effective model on the honeycomb lattice,
  in which selective random defects are represented as modulations of hopping amplitudes.
  This effective model qualitatively reproduces the topological transition within the same energy range 
  observed in the systems with selective random defects.
  This finding indicates that the topological transition induced by selective random defects can be
  interpreted as a consequence of hopping-amplitude modulations.

  The lattice structures discussed in this study can be 
  realized in a broad range of two-dimensional material platforms.
  The BK lattice structure has been identified in a quenched van der Waals
  ferromagnet Fe$_{5-\delta}$GeTe$_{2}$ \cite{Wu2024, A.F.May2019},
  vacancy-engineered graphene \cite{PhysRevB.105.155414, PhysRevB.105.014511},
  and Kekulé graphene with bond order \cite{NatPhys.10.950, PhysRevB.106.245116}.
  In addition to these systems, promising candidates for the honeycomb lattice with selective random defects 
  include van der Waals materials and its layered systems \cite{vdW2_2013, vdW_layer_rev_2017},
  graphene nanostructures \cite{Bai2010, Zhang2017},
  and photonic crystals \cite{Lu2014, PhysRevB.80.155103}.
  Furthermore, metal-organic frameworks (MOF) offer a highly tunable
  architecture that could serve as an ideal platform to implement selective random defects \cite{advs.201900506}.
  Our findings suggest that such defects can effectively modulate the underlying parameters
  in materials, thereby enabling control over their spectral and even topological properties.

  In two-dimensional electronic systems, randomness can give rise to localization effects \cite{PhysRev.109.1492},
  which are often identified by measures such as the inverse participation ratio \cite{RevModPhys.80.1355}.
  This phenomenon in two-dimensional systems is highly subtle and requires careful and systematic analysis,
  which we leave for future work.
  In addition, while the present analysis was based on a single-particle model of spinless fermions,
  future studies incorporating many-body interactions and additional degrees of freedom,
  such as spin and orbital, may reveal a variety of novel quantum phenomena.
  Finally, beyond the triangular-lattice motif in Fig.~\ref{fig:lattice},
  extending the present approach to other lattice pairs connected by selective defects, 
  such as the square and Lieb lattices
  or the three-dimensional face-centered cubic and pyrochlore lattices,
  could uncover much richer quantum phenomena.
  These may include topological transitions driven by selective random defects,
  the breakdown of topology, and the realization of topological Anderson insulators  \cite{PhysRevLett.102.136806, PhysRevB.80.165316, PhysRevLett.103.196805}.
  Exploring the interplay between quantum degrees of freedoms and selective randomness
  could open new pathways in the design and control of quantum materials.

\section*{\label{sec5}Acknowledgments}
This work was supported by the JSPS KAKENHI (Grants No.~JP22K03509, No.~JP22K13998, No.~JP23K25816, No.~JP24K17009, and No.~JP25H01247).
S.I. was supported by the
Program for Leading Graduate Schools (MERIT-WINGS) and
Hirose Foundation.

\appendix
\section{\label{Appendix}Topological transition at another parameter}

  \begin{figure}[t]
    \hypertarget{fig:App1}{}
    \includegraphics[width=\columnwidth]{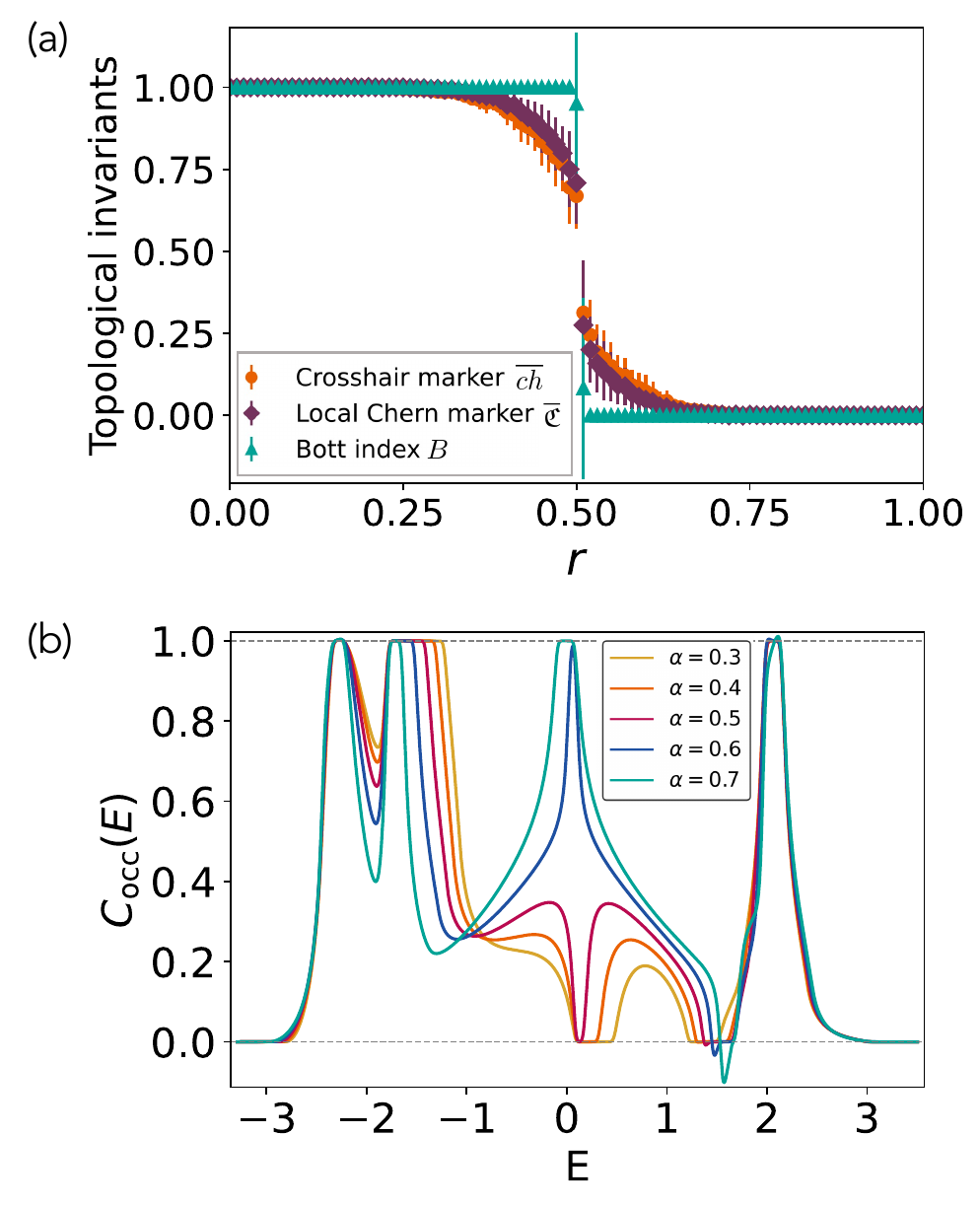}
    \caption{\label{fig:App1} 
    (a) $r$ dependencies of three topological invariants of the Haldane model on the honeycomb lattice with selective random defects.
    (b) Energy dependencies of the cumulative Berry curvature of the effective model in Eq.~\eqref{eq:Heff} at $\alpha = 0.3$, $0.4$, $0.5$, $0.6$, and $0.7$.
    The model parameters are $M/t_2=-3.7, t_2=0.3$, and $\phi=\pi/2$ for both panels.
    }
  \end{figure}

  In this Appendix, we present additional results demonstrating the correspondence between
  the topological transition induced by selective random defects with defect ratio $r$ and 
  that in the effective model with parameter $\alpha$.
  In the main text, we showed the results for $M/t_2=-3$, $t_2=0.3$, and $\phi=\pi/2$,
  where the topological transition induced by selective random defects occurs at $r\simeq0.7$,
  while that in the effective model takes place at $\alpha\simeq0.3$.
  Here, we present another example at $M/t_2=-3.7$, $t_2=0.3$, and $\phi=\pi/2$.
  Figure \hyperlink{fig:App1}{\ref{fig:App1}(a)} shows $r$ dependence of three topological invariants calculated for the model in Eq.~\eqref{eq:Hamiltonian1}.
  The chemical potential is set in the middle of the relevant gap, as we did in Fig.~\ref{fig:marker}.
  We observe a clear jump in all topological invariants at $r\simeq0.5$,
  indicating a topological transition induced by selective random defects.
  For comparison, we calculate the energy dependence of the cumulative Berry curvature $C_{\text{occ}}$ defined in Eq.~\eqref{eq:Cocc}
  for the effective model in Eq.~\eqref{eq:Heff} with the same parameter set, as shown in Fig.~\ref{fig:App1}(b).
  We find that the topological transition occurs at $\alpha\simeq0.5$,
  where the Dirac node appears at the $\Gamma$ point in the band structure (not shown).
  Thus, this additional analysis further supports the relation $\alpha \simeq 1 - r$.


\bibliography{reference}

\end{document}